\documentclass[12pt]{article}

\usepackage{cite}
\usepackage{amsmath}
\usepackage{amssymb}

\def\para{\\ [-2mm]}
\def \be  {\begin{equation}}
\def \ee  {\end{equation}}
\def \ba  {\begin{eqnarray}}
\def \ea  {\end{eqnarray}}
\newcommand{\nn}{\nonumber}
\def\eqn#1{eq.~(\ref{#1})} 
\def\Eqn#1{Equation~(\ref{#1})}
\def\eqns#1#2{eqs.~(\ref{#1}) and~(\ref{#2})}
\def\Eqns#1#2{Eqs.~(\ref{#1}) and~(\ref{#2})}

\def\IZ{\relax\ifmmode\mathchoice
{\hbox{\cmss Z\kern-.4em Z}}{\hbox{\cmss Z\kern-.4em Z}}
{\lower.4pt\hbox{\cmsss Z\kern-.4em Z}}
{\lower1.2pt\hbox{\cmsss Z\kern-.4em Z}}\else{\cmss Z\kern-.4em Z}\fi}
\newcommand{\Z}{\mathsf{Z}\kern -5pt \mathsf{Z}}
\newcommand{\unit}{\mathsf{1}\kern -3pt \mathsf{l}}

\def \Tr {\mathop{\rm Tr}\nolimits}

\def\half{{\textstyle{1 \over 2}}}
\def\fr#1#2{ {\textstyle{#1 \over #2}}}

\def\veps{\varepsilon}
\def\cA {  {\cal A} }

\def\cAn {  {\cal A}_n }

\def\cL {  {\cal L}  }

\def\cO {  {\cal O} }
\def\< { \langle}
\def\> { \rangle}

\def\ta {\textsf{a}}
\def\tb {\textsf{b}}
\def\tc {\textsf{c}}
\def\td {\textsf{d}}
\def\te {\textsf{e}}
\def\ti {\textsf{i}}
\def\tj {\textsf{j}}
\def\tk {\textsf{k}}
\def\tf{\tilde{f}}
\def\tN{\tilde{N}}
\def\bc{  {\bf c} }
\def\Sai{ S_{a,i} }
\def\lam{\lambda}

\begin{document}

\titlepage
\begin{flushright}
BOW-PH-171
\end{flushright}

\vspace{3mm}

\begin{center}
{\Large\bf\sf
Color-factor symmetry of the amplitudes  \\
of Yang-Mills and biadjoint scalar theory \\[2mm]
using perturbiner methods
}

\vskip 1.5cm

{\sc
Stephen G. Naculich$^{a}$
}

\vskip 0.5cm
$^a${\it
Department of Physics and Astronomy\\
Bowdoin College\\
Brunswick, ME 04011 USA
}

\vspace{5mm}

{\tt
naculich@bowdoin.edu
}
\end{center}

\vskip 1.5cm

\begin{abstract}
Color-factor symmetry is a property of 
tree-level gauge-theory amplitudes containing
at least one gluon. 
BCJ relations among color-ordered amplitudes 
follow directly from this symmetry.
Color-factor symmetry is also a 
feature of biadjoint scalar theory amplitudes
as well as of their equations of motion.
In this paper, we present a new proof of 
color-factor symmetry using a recursive method 
derived from the perturbiner expansion of the classical
equations of motion.

\end{abstract}

\vspace*{0.5cm}
\vfil\break

\section{Introduction}
\setcounter{equation}{0}
\label{sec:intro}

The discovery of {\it color-kinematic duality} 
in the amplitudes of Yang-Mills theory,
and subsequently in the amplitudes 
of a much broader class of field theories 
(see ref.~\cite{Bern:2019prr} for a review),
has unleashed a tool of great power, 
particularly in the calculation of gravitational amplitudes 
through the double-copy 
procedure \cite{Bern:2008qj,Bern:2010ue,Bern:2010yg}.
In 2008, Bern, Carrasco, and Johansson (BCJ) 
showed that the assumption of color-kinematic duality in 
tree-level amplitudes of Yang-Mills theory 
implies a set of linear relations among the color-ordered amplitudes.
The subsequent proof of these BCJ relations 
using string-theory techniques
\cite{BjerrumBohr:2009rd,Stieberger:2009hq}
and BCFW on-shell recursion \cite{Feng:2010my,Chen:2011jxa}
provided evidence for the conjecture of tree-level color-kinematic duality.
Bern et al. also conjectured that color-kinematic duality 
applies to integrands of loop-level amplitudes \cite{Bern:2008qj,Bern:2010ue};
while not proven, this conjecture has been tested 
for amplitudes of various multiplicities and loop levels 
in supersymmetric Yang-Mills theories,
which have been used to construct supergravity amplitudes \cite{Bern:2019prr}.
\para

In 2016, R.~W.~Brown and the current author observed that 
tree-level gauge-theory amplitudes possess a {\it color-factor symmetry},  
which acts as a momentum-dependent shift on the 
color factors of an amplitude, 
leaving the full amplitude 
invariant \cite{Brown:2016mrh,Brown:2016hck,Brown:2018wss}.
This symmetry was proved for both Yang-Mills theory 
and for gauge theories with massive particles of various spins
using the {radiation vertex expansion} \cite{Brown:1982xx}.
The BCJ relations follow as an immediate consequence of color-factor 
symmetry \cite{Brown:2016mrh,Brown:2016hck}.
\para

Color-kinematic duality 
and color-factor symmetry are 
closely related features of gauge theories: 
the former implies the latter 
(as proved using the {cubic vertex expansion}), 
but the latter implies a less stringent (but gauge-invariant)
constraint than the former 
on the kinematic numerators of a tree-level amplitude \cite{Brown:2016mrh}.
Similarly color-kinematic duality implies color-factor
symmetry for loop-level amplitudes, but no independent proof of
the latter has yet been developed.
\para

Color-factor symmetry is also a property of tree-level amplitudes 
of the biadjoint scalar (BAS) theory \cite{Cachazo:2013iea}, 
whose fields transform in the adjoint representation of $U(N) \times U(\tN)$,
as was proved using the cubic vertex expansion \cite{Brown:2016mrh}.
Cheung and Mangan \cite{Cheung:2021zvb} observed that the 
classical equations of motion of the BAS theory 
also possess color-factor symmetry, 
and that this implies the invariance of the tree-level amplitudes.
They demonstrated a relation between the 
$U(N)$ color-factor symmetry of the equations of motion
and the conservation of current associated with 
the dual $U(\tN)$ symmetry.
In ref.~\cite{Cheung:2022pdk}, these results were 
generalized to curved symmetric spacetime.
\para

In this paper, we offer a new proof of color-factor symmetry 
based on a recursive approach.
In 1987, Berends and Giele \cite{Berends:1987me}
introduced a method for computing tree-level QCD amplitudes
using a set of partially off-shell amplitudes
(subsequently known as Berends-Giele currents), 
which were then computed recursively.
Rosly and Selivanov \cite{Rosly:1996vr,Selivanov:1998hn,Selivanov:1999as}
later showed that a perturbative solution 
of the classical equations of motion
(dubbed the {\it perturbiner expansion})
acts as a generating function for Berends-Giele currents.
Mafra, Schlotterer, 
et al.\cite{Lee:2015upy,Mafra:2015vca,Mafra:2016ltu,Garozzo:2018uzj}
also used classical equations of motion to generate Berends-Giele currents
in various theories.
Mizera and Skrzypek \cite{Mizera:2018jbh}
introduced the {\it color-dressed perturbiner expansion},
which, as we will see in this paper,
is well adapted for the demonstration of 
color-factor symmetry of tree-level amplitudes. 
\para

Further developments in this subject include
the work of Lopez Arcos, Quintero V\'elez, et. al.,
who related the $L_\infty$-algebra 
that appears in Batalin-Vilkovisky quantization \cite{Macrelli:2019afx}
to the perturbiner expansion 
for biadjoint scalar and Yang-Mills theories \cite{Lopez-Arcos:2019hvg}
as well as in gauge theories with matter \cite{Gomez:2020vat}.
Berends-Giele currents in BCJ gauge were constructed 
using Bern-Kosower rules \cite{Ahmadiniaz:2021fey},
with this work extended to gravity using the double-copy 
procedure \cite{Ahmadiniaz:2021ayd}.
Gomez and Jusinskas have applied perturbiner methods 
to gravity coupled to matter \cite{Gomez:2021shh}, 
and in ref.~\cite{Armstrong:2022mfr}, perturbiner methods were used 
to compute tree-level boundary correlators in anti-de Sitter space.
The perturbiner approach has also been found effective for computing
one-loop integrands \cite{Gomez:2022dzk}.
The connection between tree-level Berends-Giele recursion relations 
and the $L_\infty$-algebra uncovered in ref.~\cite{Macrelli:2019afx}
was extended to loop-level recursion relations and 
the quantum homotopy algebra $A_\infty$ in ref.~\cite{Jurco:2019yfd}. 
For connections between the homotopy algebra and the double copy,
see refs.~\cite{Borsten:2021hua,Escudero:2022zdz,Borsten:2022ouu}.
\para

We present this alternative proof of color-factor symmetry
because the recursive methods employed 
may be more familiar to modern readers
than the radiation vertex expansion used in 
ref.~\cite{Brown:2016mrh} to prove color-factor symmetry.
Moreover, 
this recursive approach may be easier to generalize
to the exploration of color-factor symmetry in other theories.
\para

The outline of this paper is as follows.
In sec.~\ref{sec:cfs}, 
we recall how color-factor symmetry acts on amplitudes,
and how the BCJ relations follow as a consequence.
In sec.~\ref{sec:perturbiner}, 
we show how the color-dressed perturbiner expansion
is used to compute tree-level amplitudes,
first in the biadjoint scalar theory,
and then in Yang-Mills theory.
In sec.~\ref{sec:proof},  we then use the color-dressed perturbiner
expansion to prove color-factor symmetry 
for tree-level amplitudes of the biadjoint scalar theory
and of Yang-Mills theory.
Sec.~\ref{sec:concl} contains our conclusions.

\section{Color-factor symmetry and BCJ relations}
\setcounter{equation}{0}
\label{sec:cfs}

Tree-level scattering amplitudes of a gauge theory 
are given by a sum of Feynman diagrams, 
and can be expressed as \cite{Bern:2010tq}
\begin{align}
\cA_n =\sum_i a_i c_i
\label{colordecomposition}
\end{align}

where $c_i$ are {\it color factors},
consisting of the contraction of various color tensors
$f^{\ta\tb\tc}$ and $\left( T^{\ta} \right)^\ti_{~\tj} $
appearing in the Feynman diagrams,
and $a_i$ depends on kinematic and spin factors.
Each color factor can itself be represented as a Feynman 
diagram \cite{Cvitanovic:1976am,Cvitanovic:1980bu},
one that contains only trivalent vertices.
(If the full Feynman diagram contains only trivalent vertices, 
then it contributes to the color factor with the same Feynman diagram.
If the full Feynman diagram also contains quartic vertices,
then its contribution is parcelled out among different color factors
by expressing the quartic vertex as products of trivalent vertices.) 
Note that, due to the group theory identities 
\begin{align}
0
&=
 f^{\tb\ta\te} f^{\te \tc \td}
+f^{\tc\ta\te} f^{\te \td \tb} 
+f^{\td\ta\te} f^{\te \tb \tc} \,,
\label{jacobi}\\[3mm]
0
&=
  \left( T^{\ta } \right)^\ti_{~\tk}  \left( T^{\tc } \right)^\tk_{~\tj}  
- \left( T^{\tc } \right)^\ti_{~\tk}  \left( T^{\ta } \right)^\tk_{~\tj}  
- f^{\ta \tc \te} 
  \left( T^{\te } \right)^\ti_{~\tj}  
\label{liealgebra}
\end{align}
there exist (Jacobi) relations among the various color factors $c_i$.
Since the $c_i$ are not independent, 
there is some choice about how the coefficients $a_i$ are defined.

\subsection{Color-factor symmetry}

There exists a color-factor symmetry associated with each external 
gluon $a$ contributing to the amplitude \cite{Brown:2016mrh,Brown:2016hck}. 
This symmetry acts on each color factor $c_i$ 
appearing in \eqn{colordecomposition} by a 
momentum-dependent shift $\delta_a c_i$.
For each color factor $c_i$,
the gluon leg $a$ divides the associated tree-level diagram in two 
at its point of attachment.
Let $\Sai$ denote the subset of the remaining legs on one side
of this point; it does not matter which side we choose.
The shift of the color factor $c_i$ associated with gluon $a$
then satisfies\footnote{Choosing to sum 
over the complement of $\Sai$ gives the same result (up to sign)
due to momentum conservation.}
\begin{align}
\delta_a c_i  ~\propto~  \sum_{d\in \Sai  } k_a \cdot k_d
\label{deltac}
\end{align}
where $k_a^\mu$ is the outgoing momentum of gluon $a$ (satisfying $k_a^2=0$),
and $k_d^\mu$ are the outgoing momenta of the legs belonging to $\Sai$.
The color-factor shift also respects the group theory identities,
as we will see below.
\para

We may regard the color-factor symmetry 
as acting directly on the color tensors appearing in $c_i$.
If gluon $a$ (with color $\ta$) is attached to a gluon line,
so that the color factor contains $f^{\tb \ta \tc}$,
then the color-factor symmetry acts as \cite{Cheung:2021zvb}
\begin{align}
\delta_a f^{\tb \ta \tc} =  \alpha_a \delta^{\tb\tc} (k_c^2 - k_b^2)
\label{cfs}
\end{align}
where $k^\mu_b$ and $k^\mu_c$ are the momenta 
flowing out of the vertex associated with $f^{\tb \ta \tc} $
and $\alpha_a$ is a constant parameter.
If gluon $a$ is attached to a line 
corresponding to a particle in some other representation,
so that the color factor contains 
$\left( T^{\ta } \right)^\ti_{~\tj}  $,
then the color-factor symmetry acts as 
\begin{align}
\delta_a \left( T^{\ta } \right)^\ti_{~\tj}  
=  \alpha_a 
\delta^{\ti}_{~\tj} 
(k_j^2 - k_i^2)
\end{align}
where $k^\mu_i$ and $k^\mu_j$ are the momenta 
flowing out of the vertex associated with 
$\left( T^{\ta } \right)^\ti_{~\tj}  $.
Using momentum conservation at each vertex,
we may express these shifts as 
\begin{align}
\delta_a f^{\tb \ta \tc} =   \alpha_a \delta^{\tb\tc} (2 k_a \cdot k_b ) \,,
\qquad \qquad
\delta_a \left( T^{\ta } \right)^\ti_{~\tj}  
=  \alpha_a 
\delta^{\ti}_{~\tj} 
(2 k_a \cdot k_i)  \,.
\label{deltaf}
\end{align}
The relations (\ref{deltaf}) guarantee 
that the color-factor shifts satisfy \eqn{deltac}.
We must also check that the color-factor shifts 
leave the group theory identities invariant.
Using \eqn{deltaf} in \eqns{jacobi}{liealgebra}, we find
\begin{align}
&
\delta_a \left[ 
 f^{\tb\ta\te} f^{\te \tc \td}
+f^{\tc\ta\te} f^{\te \td \tb}
+f^{\td\ta\te} f^{\te \tb \tc}
\right]
= 2 \alpha_a k_a \cdot (k_b + k_c + k_d) f^{\tb\tc\td}  
= - 2 \alpha_a k_a^2 f^{\tb\tc\td}   =0 \,,
\\[3mm]
&
\delta_a 
  \left[
\left( T^{\ta } \right)^\ti_{~\tk}  \left( T^{\tc } \right)^\tk_{~\tj}  
- \left( T^{\tc } \right)^\ti_{~\tk}  \left( T^{\ta } \right)^\tk_{~\tj}  
- f^{\ta \tc \te} \left( T^{\te } \right)^\ti_{~\tj}  
\right]
=
2 \alpha_a k_a \cdot (k_i + k_j + k_c ) 
 \left( T^{\tc } \right)^\ti_{~\tj}  
=
-2  \alpha_a k_a^2  
 \left( T^{\tc } \right)^\ti_{~\tj}   =0
\nn
\end{align}
using momentum conservation and the masslessness of the gluon.
\para

In ref.~\cite{Brown:2016mrh},
the $n$-point amplitude \eqn{colordecomposition}
was proved to be invariant under the color-factor shift 
associated with any of the external gluons it contains
\begin{align}
\delta_a \cA_n = 0
\end{align}
by rewriting the amplitude using the radiation vertex expansion.
In sec.~\ref{sec:proof} we give an alternative proof of this fact
using the recursive perturbiner approach.

\subsection{BCJ relations}

In the remainder of this section, 
we recall the demonstration \cite{Brown:2016mrh}
that color-factor symmetry of the amplitude implies
the fundamental BCJ relation 
\cite{BjerrumBohr:2009rd,Feng:2010my,Sondergaard:2011iv}
among the color-ordered amplitudes.
As mentioned above, the color factors $c_i$ are not independent
due to group theory identities (\ref{jacobi}) and (\ref{liealgebra}).
It is useful to identity an independent basis of color factors, 
whose coefficients will be unambiguously specified \cite{Cvitanovic:1980bu}.
For tree-level $n$-gluon amplitudes, 
such a basis consists of half-ladder color factors
\begin{align}
\bc_{1 \gamma n} 
&\equiv  
\sum_{\tb_1,\ldots,\tb_{n{-}3}}
f^{\ta_1 \ta_{\gamma(2)} \tb_1}
f^{\tb_1 \ta_{\gamma(3)} \tb_2} \cdots 
f^{\tb_{n{-}3} \ta_{\gamma(n{-}1)} \ta_n} 
\label{halfladdergamma}
\end{align}
in terms of which the amplitude may be written 
as \cite{DelDuca:1999ha,DelDuca:1999rs}
\begin{align}
\cAn ~=~ \sum_{\gamma \in S_{n-2}}
\bc_{1 \gamma n } A(1, \gamma(2), \cdots, \gamma(n-1), n)
\label{ddm}
\end{align}
where $\gamma$ runs over all permutations of $\{2, \cdots, n-1\}$, 
and $A(1, \gamma(2), \cdots, \gamma(n-1), n)$ are color-ordered amplitudes.
Singling out one of the external gluons ($a=2$)
and letting $\sigma$ denote an arbitrary permutation of $\{3, \cdots, n-1\}$,
we may reexpress \eqn{ddm} as
\begin{align}
\cAn &= \sum_{\sigma \in S_{n-3}}
\left[ \sum_{e=3}^{n} 
\bc_{1 \sigma(3) \cdots \sigma(e-1) 2 \sigma(e) \cdots  \sigma(n-1) n }
A(1, \sigma(3), \cdots, \sigma(e-1), 2, \sigma(e), \cdots, \sigma(n-1), n)
\right]
\end{align}

where 
\begin{align}
&
\bc_{1 \sigma(3) \cdots \sigma(e-1) 2 \sigma(e) \cdots  \sigma(n-1) n }
\nn\\
&
\qquad
= \sum_{\tb_1,\ldots,\tb_{n{-}3}}
f^{\ta_1 \ta_{\sigma(3)} \tb_{1}  }
\cdots
f^{\tb_{e-4} \ta_{\sigma(e-1)} \tb_{e-3}  }
f^{\tb_{e-3} \ta_2  \tb_{e-2} }
f^{\tb_{e-2} \ta_{\sigma(e)} \tb_{e-1}} 
\cdots 
f^{\tb_{n-3}  \ta_{\sigma(n-1)} \ta_{n}  } \,.
\label{halfladdersigma}
\end{align}

The color-factor symmetry associated with gluon $a=2$
acts on \eqn{halfladdersigma} as
\begin{align}
\delta_2 \, 
\bc_{1 \sigma(3) \cdots \sigma(e-1) 2 \sigma(e) \cdots  \sigma(n-1) n }
&= 
2 \alpha_2 k_2 \cdot \left(
k_1  +  \sum_{d=3}^{e-1} k_{\sigma(d)} \right)
\bc_{1 \sigma(3) \cdots \sigma(e-1) \sigma(e) \cdots  \sigma(n-1) n }
\end{align}

and therefore 
\begin{align}
\delta_2 \, \cAn &= 2\alpha_2 
\sum_{\sigma  \in S_{n-3} } 
\bc_{1 \sigma n}
\sum_{e=3}^n k_2 \cdot \left( k_1 + \sum_{d=3}^{e-1} k_{\sigma(d)}  \right)
A(1, \sigma(3), \cdots, \sigma(e-1), 2, \sigma(e), \cdots, \sigma(n-1), n)  \,.
\end{align}
Since $\delta_2 \cA_n=0$ by color-factor symmetry, 
and since the half-ladder color factors 
$\bc_{1 \sigma n}$ are independent, 
this establishes that
\be
\sum_{e=3}^n \left( k_2 \cdot k_1 + \sum_{d=3}^{e-1} k_2 \cdot k_{\sigma(d)}  \right)
A(1, \sigma(3), \cdots, \sigma(e-1), 2, \sigma(e), \cdots, \sigma(n-1), n)
~=~ 0
\label{fundbcj}
\ee
which is the fundamental BCJ relation,
from which the rest of the BCJ relations may be 
derived \cite{BjerrumBohr:2009rd,Feng:2010my,Sondergaard:2011iv}.
This argument may be generalized to the amplitudes of the BAS theory
in curved symmetric spacetime \cite{Cheung:2022pdk}.
\para

For tree-level amplitudes containing fields in other representations 
(e.g. quarks) in addition to gluons, 
an independent basis of color factors
is given by the Melia basis \cite{Melia:2013bta,Melia:2013epa,Melia:2015ika}.
The independent amplitudes corresponding to this basis
also satisfy BCJ relations that follow from the assumption
of color-kinematic duality \cite{Johansson:2015oia,delaCruz:2015dpa}. 
Color-factor symmetry can also be used 
to derive BCJ relations for these amplitudes \cite{Brown:2016hck}.

\section{Color-dressed perturbiner expansion}
\setcounter{equation}{0}
\label{sec:perturbiner}

In this section, we review the 
color-dressed perturbiner expansion \cite{Mizera:2018jbh}
of the solutions to the classical equations of motion
for the biadjoint scalar theory and Yang-Mills theory,
and how its coefficients (Berends-Giele currents) 
are used to obtain tree-level $n$-point amplitudes in those theories.

\subsection{Biadjoint scalar theory}

The biadjoint scalar theory
is a theory of a massless scalar field $\phi^{\ta\ta'}$ 
transforming in the adjoint representation of $U(N) \times U(\tN)$,
with Lagrangian \cite{Cachazo:2013iea}
\begin{align}
\cL = \half 
\big( \partial_\mu \phi^{\ta \ta'} \big)
\big( \partial^\mu \phi^{\ta \ta'} \big)
~-~ 
\fr{1}{6} \lam
f^{\ta\tb\tc} \tf^{\ta' \tb' \tc'} 
\phi^{\ta \ta'} \phi^{\tb \tb'} \phi^{\tc \tc'} 
\end{align}

where $f^{\ta\tb\tc} $ and $\tf^{\ta' \tb' \tc'}$ are the structure 
constants\footnote{Normalized by 
$f^{\ta \tb \tc } = \Tr( [T^\ta, T^\tb] T^\tc )$
with $\Tr( T^\ta T^\tb ) = \delta^{\ta\tb}$,
so that $[T^\ta, T^\tb] = f^{\ta\tb\tc} T^\tc$.
We use $\eta_{00}=1$.
\label{ftnt:struc}}
of $U(N)$ and $U(\tN)$ respectively.
This Lagrangian yields the equation of motion 
\be
\partial^2 \phi^{\ta \ta'}
=
~-
\half \lam f^{\ta\tb\tc} \tf^{\ta' \tb' \tc'}  
\phi^{\tb \tb'} \phi^{\tc \tc'}  \,.
\label{BASeom}
\ee

Rosly and Selivanov \cite{Rosly:1996vr,Selivanov:1998hn,Selivanov:1999as}
introduced the {\it perturbiner ansatz},
which is a solution to the nonlinear classical equation of motion
obtained by first solving the free equation of motion
$\partial^2 \phi^{\ta \ta'} = 0$
with an arbitrary linear combination of plane waves\footnote{The number
$M$ of plane waves is arbitrary, but the light-like momenta $k^\mu_i$ will 
eventually be taken as the momenta of external states in an 
$n$-point amplitude, so $M$ should at least equal the multiplicity.}
\begin{align}
\phi^{\ta\ta'}   (x) 
&= \sum_{i=1}^M \phi_i^{\ta\ta'} e^{ik_i \cdot x}  + \cO(\lam)\,,
\qquad
\phi_i^{\ta\ta'} = \veps_i \delta^{\ta \ta_i} \delta^{\ta' \ta'_i}  \,,
\qquad
k_i^2=0
\label{BASplane}
\end{align}

and then using this as a seed in \eqn{BASeom} 
to generate corrections higher order in $\lambda$.
The coefficients of this expansion are used 
to compute tree-level amplitudes of the theory. 
\para

For the purpose of proving 
the color-factor symmetry of tree-level amplitudes in sec.~\ref{sec:proof}, 
we find it convenient to use a version 
of the perturbiner ansatz
developed by Mizera and Skrzypek \cite{Mizera:2018jbh},
called the {\it color-dressed perturbiner expansion}
(in distinction from the color-stripped perturbiner expansion).
For the BAS theory, the ansatz can be 
written\footnote{In the higher order terms, 
one suppresses terms in which some of the indices
$i$, $j$, $k$ coincide. 
This may be achieved 
formally \cite{Rosly:1996vr,Selivanov:1998hn,Selivanov:1999as}
by setting $\veps_i^2=0$.}
\begin{align}
\phi^{\ta\ta'}   (x) 
&= \sum_{i} \phi_i^{\ta\ta'}\, e^{i k_i\cdot x} 
+ \sum_{i<j} \phi_{ij}^{\ta\ta'}\, e^{i k_{ij}\cdot x} 
+ \sum_{i<j<k} \phi_{ijk}^{\ta\ta'}\, e^{i k_{ijk}\cdot x} + \cdots
\label{BASperturbiner}
\end{align}
where $k_{ij}^\mu = k_i^\mu + k_j^\mu$, etc.
\Eqn{BASperturbiner} is expressed compactly as 
\begin{align}
\phi^{\ta\ta'} (x) = 
 \sum_P \phi_P^{\ta\ta'}  e^{i k_P \cdot x}
\label{BASP}
\end{align}
summing over all non-empty ordered words $P=p_1 p_2 \cdots p_m$
with $1 \le p_1 < p_2 < \cdots < p_m \le M$,
where $k_P = \sum_{j=1}^m k_{p_j}$.
By inserting \eqn{BASP} into \eqn{BASeom} one obtains \cite{Mizera:2018jbh}
\begin{align}
 \phi_P^{\ta\ta'}  
= 
{\lam \over 2 k_P^2} f^{\ta\tb\tc} \tf^{\ta' \tb' \tc'}  
\sum_{P = Q \cup R} \phi_Q^{\tb\tb'} \phi_R^{\tc\tc'}
\label{biadjointrecursion}
\end{align}

where $P = Q \cup R$ denotes all possible divisions of $P$ into 
two non-empty ordered words $Q$ and $R$.
The coefficients $\phi_{P}^{\ta\ta'}$ are Berends-Giele
currents\footnote{Berends and Giele \cite{Berends:1987me}
originally defined these currents for Yang-Mills theory,   
which Mafra \cite{Mafra:2016ltu} adapted to the BAS theory.}
of the BAS theory, computed recursively using \eqn{biadjointrecursion}.
One sees that $\phi_P^{\ta\ta'}$  has a pole at $k_P^2=0$.
\para

To obtain the tree-level $n$-point amplitude $\cA_n$,
one first computes $\phi_P^{\ta\ta'}$  for $P=12\cdots (n-1)$.
Since momentum conservation for the $n$-point amplitude 
implies $k_P = - k_n$, and an on-shell amplitude has $k_n^2=0$,
one extracts the residue of the $k_P^2$ pole
of $\phi_P^{\ta\ta'}$  and contracts with $\phi_n^{\ta\ta'}$ 
to get \cite{Mizera:2018jbh}
\begin{align}
\cA_n 
&= \lim_{k_P^2 \to 0} \phi_n^{\ta\ta'} k_P^2 \phi_P^{\ta\ta'} \,.
\label{npointscalar}
\end{align}

To illustrate this procedure for the four-point amplitude
we first use \eqns{BASplane}{biadjointrecursion} to compute the 
rank-2 perturbiner coefficient
\begin{align}
\phi_{ij}^{\ta\ta'} 
&= 
{\lam \over  2k_{ij}^2} f^{\ta\tb\tc} \tf^{\ta' \tb' \tc'}  
\left( 
 \phi_i^{\tb\tb'} \phi_j^{\tc\tc'}
 +\phi_j^{\tb\tb'} \phi_i^{\tc\tc'}
\right)
= 
{ \lam \veps_i \veps_j \over  k_{ij}^2} f^{\ta\ta_i\ta_j} \tf^{\ta' \ta_i' \ta_j'}  
\end{align}

and from this the rank-3 coefficient 
\begin{align}
\phi_{123}^{\ta\ta'} 
&= 
{\lam \over 2 k_{123}^2} f^{\ta\tb\tc} \tf^{\ta' \tb' \tc'}  
\left( 
\phi_{12}^{\tb\tb'} \phi_{3}^{\tc\tc'}
+\phi_{3}^{\tb\tb'} \phi_{12}^{\tc\tc'}
+\phi_{13}^{\tb\tb'} \phi_{2}^{\tc\tc'}
+\phi_{2}^{\tb\tb'} \phi_{13}^{\tc\tc'}
+\phi_{23}^{\tb\tb'} \phi_{1}^{\tc\tc'}
+\phi_{1}^{\tb\tb'} \phi_{23}^{\tc\tc'}
\right)
\nn\\
&= 
{\lam^2 \veps_1 \veps_2 \veps_3 \over k_{123}^2} 
\left[ 
{ 
c^\ta_{123} \tilde{c}^{\ta'}_{123} \over k_{12}^2 }
+ \hbox{(cyclic permutations of 123)} \right]
\end{align}
where 
$c^\ta_{123}  = f^{\ta_1\ta_2\tc} f^{\tc\ta_3\ta} $
and 
$\tilde{c}^{\ta'}_{123}  = \tf^{\ta'_1\ta'_2\tc'} \tf^{\tc'\ta'_3\ta'} $.
Then \eqn{npointscalar} is used
to obtain the four-point amplitude 
(setting $\veps_1 \veps_2 \veps_3 \veps_4=1$)
\begin{align}
\cA_4 = 
\lam^2
\left[ 
{
c_{1234} \tilde{c}_{1234} \over k_{12}^2 }
+ \hbox{(cyclic permutations of 123)} \right]
\end{align}
where 
$c_{1234}  = f^{\ta_1\ta_2\tc} f^{\tc\ta_3\ta_4} $
and 
$\tilde{c}_{1234}  = \tf^{\ta'_1\ta'_2\tc'} \tf^{\tc'\ta'_3\ta_4'} $.
This result agrees with four-point amplitude found in ref.~\cite{Cachazo:2013iea}.

\subsection{Yang-Mills theory}

We now describe the color-dressed perturbiner expansion for Yang-Mills 
theory \cite{Mizera:2018jbh}.
The Yang-Mills Lagrangian
\begin{align}
\cL = -\fr{1}{4} F_{\mu\nu}^\ta F^{\mu\nu\,\ta}
\end{align}

implies the equation of motion\footnote{See footnote \ref{ftnt:struc}.}
\begin{align}
\partial_\nu F^{\nu\mu\,\ta} = ig f^{\ta\tb\tc} A_\nu^\tb F^{\nu\mu\,\tc} 
\label{YMeom}
\end{align}

where the Yang-Mills field strength is given by 
\begin{align}
F_{\mu\nu}^\ta = 
\partial_\mu A_\nu^\ta
-\partial_\nu A_\mu^\ta
-ig f^{\ta\tb\tc} A_\mu^\tb A_\nu^\tc \,.
\end{align}

Choosing Lorenz gauge 
\begin{align}
\partial_\nu A^{\nu \,\ta} = 0
\label{lorenz}
\end{align}

we can write \eqn{YMeom} as
\begin{align}
\partial^2 A^{\mu \,\ta} =  ig f^{\ta\tb\tc} A_\nu^\tb 
\left( \partial^\nu A^{\mu\,\tc} + F^{\nu\mu\,\tc} \right) \,.
\label{YMeombis}
\end{align}

For convenience we define
$G^{\nu\mu\ta} \equiv 
-i \left( \partial^\nu A^{\mu\,\ta}  + F^{\nu\mu\,\ta} \right)$,
which becomes 
\begin{align}
G^{\nu\mu\ta} 
&= 
-i \left( 2 \partial^\nu A^{\mu\,\ta}  - \partial^\mu A^{\nu\,\ta}   \right)
- g f^{\ta\tb\tc} A^{\nu\, \tb} A^{\mu\,\tc}
\label{defG}
\end{align}

so that \eqn{YMeombis} is expressed as 
\begin{align}
\partial^2 A^{\mu\, \ta} &=  -g f^{\ta\tb\tc} A_\nu^\tb G^{\nu\mu\,\tc}  \,.
\label{Aeom}
\end{align}

The advantage of using 
two fields, $A^{\mu \ta}$ and $G^{\nu\mu \ta}$,
rather than just $A^{\mu\ta}$
is that \eqns{defG}{Aeom} contain only quadratic (not cubic) terms,
simplifying the recursion relations derived 
below \cite{Lee:2015upy,Mafra:2015vca,Mafra:2016ltu,Garozzo:2018uzj,Mizera:2018jbh}.
\para

We now solve these equations with the perturbiner ansatz.
As in the previous subsection, we begin by solving the free equation 
\begin{align}
\partial^2 A^{\mu\,\ta} =0, \qquad\qquad \partial_\nu A^{\nu \,\ta} = 0
\end{align}

with an arbitrary linear combination of plane waves 
\begin{align}
A^{\mu \,\ta}(x)  &= \sum_{i=1}^M A_i^{\mu \,\ta} e^{ik_i \cdot x}
+ \cO(g)
\qquad\hbox{with}\qquad k_i^2=0
\end{align}

where
\begin{align}
A_i^{\mu \,\ta}  &= \veps_i^\mu \delta^{\ta\ta_i} 
\qquad\hbox{with}\qquad \veps_i \cdot k_i = 0 \,.
\label{defAi}
\end{align}

Also to this order we have 
\begin{align}
G^{\nu\mu \,\ta}(x)  &= \sum_{i=1}^M G_i^{\nu\mu \,\ta} e^{ik_i \cdot x}
+ \cO(g)
\end{align}

where
\begin{align}
G_i^{\nu\mu\,\ta} &= g_i^{\nu\mu} \delta^{\ta\ta_i} 
\qquad\hbox{with}\qquad 
g_i^{\nu\mu} = 2 k_i^\nu \veps_i^\mu - k_i^\mu \veps_i^\nu \,.
\label{defGi}
\end{align}

As before, 
the lowest-order solution is the first term of the color-dressed perturbiner 
expansion\footnote{In refs.~\cite{Lee:2015upy,Mafra:2015vca,Mafra:2016ltu,Garozzo:2018uzj,Mizera:2018jbh}
the plane wave factor is written $e^{k_P \cdot x}$,
with the momentum taken imaginary,
in order to avoid a proliferation of factors of $i$. 
With the conventions of this paper, it is simpler  to write
$e^{ik_P \cdot x}$ 
and use real momenta.}
\begin{align}
A^{\mu\,\ta} (x) = \sum_P A_P^{\mu\,\ta}  e^{i k_P \cdot x}, \qquad\qquad
G^{\nu\mu\,\ta} (x) = \sum_P G_P^{\nu\mu\,\ta} e^{i k_P \cdot x} \,.
\label{YMperturbiner}
\end{align}

The coefficients $A^{\mu \,\ta}_P$ are the 
(color-dressed) Berends-Giele currents of 
the Yang-Mills theory \cite{Berends:1987me}.
To obtain the tree-level $n$-gluon amplitude,
one first computes $A_P^{\mu\ta}$ for $P=12\cdots (n-1)$,
then extracts the residue of the $k_P^2$ pole,
and finally contracts\footnote{For notational clarity, 
we resort here and below to the regrettable practice 
of writing all Lorentz indices upstairs.
Repeated indices are of course contracted with the Minkowski metric.}
with the Berends-Giele current $A^{\mu \,\ta}_n$
of the last gluon \cite{Berends:1987me}
\begin{align}
\cA_n 
&= \lim_{k_P^2 \to 0} 
A_n^{\mu\,\ta}  k_P^2 A_P^{\mu\,\ta}  \,.
\label{ngluon}
\end{align}

The recursion relations for the Berends-Giele currents $A_P^{\mu\ta}$
are obtained by plugging \eqn{YMperturbiner} into \eqn{Aeom}
to obtain
\begin{align}
k_P^2 A_P^{\mu \,\ta} &= 
g f^{\ta\tb\tc} \sum_{P=Q \cup R} A_Q^{\nu\,\tb} G_R^{\nu\mu\,\tc}  \,.
\label{AeomP}
\end{align}

Similarly, plugging \eqn{YMperturbiner} into \eqn{defG} we find
\begin{align}
G_P^{\nu\mu\,\ta} &= 2 k_P^\nu A_P^{\mu\,\ta} - k_P^\mu A_P^{\nu\,\ta}
- H_P^{\nu\mu\,\ta}
\label{defGP} 
\end{align}

where
\begin{align}
H_P^{\nu\mu\,\ta}
&\equiv 
g f^{\ta\tb\tc} \sum_{P = Q \cup R} A_Q^{\nu\,\tb} A_R^{\mu\,\tc} \,.
\label{defHP}
\end{align}

We combine the three previous equations to obtain 
\begin{align}
k_P^2 G_P^{\nu\mu \,\ta} 
= g f^{\ta\tb\tc} \sum_{P=Q \cup R} 
\left[ 
2k_P^\nu A_Q^{\lam \,\tb} G_R^{\lam\mu\,\tc}
-k_P^\mu A_Q^{\lam \,\tb} G_R^{\lam\nu\,\tc}
-k_P^2 A_Q^{\nu \,\tb} A_R^{\mu\,\tc}
\right] \,.
\label{GeomP}
\end{align}


\Eqns{AeomP}{GeomP} play a key role in the proof of 
color-factor symmetry in the next section.
In the remainder of this section, 
we illustrate how they are used recursively 
to compute the four-gluon amplitude. 
We first use \eqns{AeomP}{GeomP} together with 
\eqns{defAi}{defGi} to obtain the rank-2 coefficients
\begin{align}
A_{ij}^{\mu\,\ta} 
&= 
{ g \over k_{ij}^2}  
f^{\ta\tb\tc}
\Bigl[ A_i^{\nu\,\tb} G_j^{\nu\mu\,\tc} 
+ (i \leftrightarrow j) \Bigr]
= 
{ g \over k_{ij}^2}  
f^{\ta\ta_i\ta_j}\Bigl[ \veps_i^\nu g_j^{\nu\mu} - (i \leftrightarrow j) \Bigr] \,,
\\[2mm]
G_{ij}^{\nu\mu\,\ta} 
&= 
{ g \over k_{ij}^2}  
f^{\ta\ta_i\ta_j}\left[ 
2 k_{ij}^\nu \veps_i^\lam g_j^{\lam\mu} 
- k_{ij}^\mu \veps_i^\lam g_j^{\lam\nu} 
- k^2_{ij} \veps_i^\nu \veps_j^{\nu}  
- (i \leftrightarrow j) \right] \,.
\end{align}

These are then used in \eqn{AeomP} to determine the rank-3 coefficient
\begin{align}
A_{123}^{\mu\,\ta} 
&= {g \over k_{123}^2}
f^{\ta\tb\tc}
\Big[ 
A_{12}^{\nu\,\tb} G_3^{\nu\mu\,\tc}
+A_{3}^{\nu\,\tb} G_{12}^{\nu\mu\,\tc}
+A_{13}^{\nu\,\tb} G_2^{\nu\mu\,\tc}
+A_{2}^{\nu\,\tb} G_{13}^{\nu\mu\,\tc}
+A_{23}^{\nu\,\tb} G_1^{\nu\mu\,\tc}
+A_{1}^{\nu\,\tb} G_{23}^{\nu\mu\,\tc}
\Big]
\nn\\
&= {g^2 \over k_{123}^2}
\left[ {c_{123}^\ta n_{123}^\mu \over k_{12}^2 } + \hbox{cyc(1,2,3)} \right]
\label{rankthree}
\end{align}

where
\begin{align}
c_{123}^\ta &= 
f^{\ta_1\ta_2\tc} f^{\tc\ta_3\ta} \,,
\\[2mm]
n_{123}^\mu &= \left[ 
\veps_1^\lam g_2^{\lam\nu} g_3^{\nu\mu}
- \veps_3^\nu \left( 
2k_{12}^\nu\veps_1^\lam g_2^{\lam\mu}
-k_{12}^\mu\veps_1^\lam g_2^{\lam\nu}
-k_{12}^2\veps_1^\nu \veps_2^\mu \right) \right] - (1 \leftrightarrow 2) \,.
\label{n123}
\end{align}

We may obtain a more explicit form for $n_{123}^\mu$ 
by using \eqn{defGi} in \eqn{n123} and simplifying 
\begin{align}
n_{123}^\mu &=
k_1^\mu \big[
  2 \veps_2 \cdot \veps_3 \ k_2 \cdot \veps_1 
- 2 \veps_1 \cdot \veps_3 \ k_1 \cdot \veps_2 
- 3 \veps_1 \cdot \veps_2 \ k_2 \cdot \veps_3 
-   \veps_1 \cdot \veps_2 \ k_1 \cdot \veps_3 
\big]
\nn\\
&+ 
k_2^\mu \big[
  2 \veps_2 \cdot \veps_3 \ k_2 \cdot \veps_1 
- 2 \veps_1 \cdot \veps_3 \ k_1 \cdot \veps_2 
+   \veps_1 \cdot \veps_2 \ k_2 \cdot \veps_3 
+ 3 \veps_1 \cdot \veps_2 \ k_1 \cdot \veps_3 
\big]
\nn\\
&+ 
k_3^\mu \big[
- 2 \veps_2 \cdot \veps_3 \ k_2 \cdot \veps_1 
+ 2 \veps_1 \cdot \veps_3 \ k_1 \cdot \veps_2 
+   \veps_1 \cdot \veps_2 \ k_2 \cdot \veps_3 
-   \veps_1 \cdot \veps_2 \ k_1 \cdot \veps_3 
\big]
\nn\\
&+ 
\veps_1^\mu \big[
  4 k_1 \cdot \veps_2 \ k_1 \cdot \veps_3
+ 4 k_1 \cdot \veps_2 \ k_2 \cdot \veps_3
- 2 \veps_2 \cdot \veps_3 \ k_1 \cdot k_2  \big]
\nn\\
&+ 
\veps_2^\mu \big[
- 4 k_2 \cdot \veps_1 \ k_1 \cdot \veps_3
- 4 k_2 \cdot \veps_1 \ k_2 \cdot \veps_3
+ 2 \veps_1 \cdot \veps_3 \  k_1 \cdot k_2 
 \big]
\nn\\
&+ 
\veps_3^\mu \big[
  4 k_2 \cdot \veps_1 \ k_3 \cdot \veps_2
- 4 k_3 \cdot \veps_1 \ k_1 \cdot \veps_2
- 2 \veps_1 \cdot \veps_2 \ k_2 \cdot k_3  
+ 2 \veps_1 \cdot \veps_2 \ k_1 \cdot k_3  \big] \,.
\end{align}

We observe that color factors 
appearing in 
the Berends-Giele current (\ref{rankthree}) satisfy 
the Jacobi identity
$c_{123}^\ta + c_{231}^\ta + c_{312}^\ta  =0$,
but the kinematic numerators do not 
\begin{align}
n_{123}^\mu 
+n_{231}^\mu 
+n_{312}^\mu 
&=
(k_1^\mu +k_2^\mu + k_3^\mu) 
\Big[
    \veps_1 \cdot \veps_2 (k_1-k_2) \cdot \veps_3 
+ \hbox{cyc(1,2,3)} 
\Big] \,.
\label{kinemnum}
\end{align}

Finally, \eqn{ngluon} yields the well known four-gluon amplitude
\begin{align}
\cA_4 =  g^2 
\left[ {c_{1234} n_{1234} \over k_{12}^2 } + \hbox{cyc(1,2,3)} \right],
\qquad 
c_{1234} = 
f^{\ta_1\ta_2\tc} f^{\tc\ta_3\ta_4},
\qquad
n_{1234} = n_{123}^\mu \veps_4^\mu  \,.
\label{fourgluon}
\end{align}

Both color factors and kinematic numerators in the four-gluon
amplitude satisfy the Jacobi identity
\begin{align}
c_{1234} + 
c_{2314} + 
c_{3124}  &= 0 \,,
\\[2mm]
n_{1234} 
+n_{2314} 
+n_{3124} 
&=
\veps_4 \cdot (k_1^\mu +k_2^\mu + k_3^\mu) 
\Big[
    \veps_1 \cdot \veps_2 (k_1-k_2) \cdot \veps_3 
+ \hbox{cyc(1,2,3)} 
\Big]
=0
\end{align}
because  $k_1 + k_2 + k_3 = -k_4$ and $\veps_4 \cdot k_4 =0$.

\section{Recursive proof of color-factor symmetry}
\setcounter{equation}{0}
\label{sec:proof}

In this section, we present proofs that the tree-level $n$-point amplitudes
of the BAS theory and Yang-Mills theory are invariant
under the color-factor shifts described in sec.~\ref{sec:cfs}
using the recursion relations derived from the 
color-dressed perturbiner expansions in sec.~\ref{sec:perturbiner}.

\subsection{Biadjoint scalar theory}

We begin by combining \eqn{npointscalar} 
with \eqns{BASplane}{biadjointrecursion}
to obtain the following expression for 
the tree-level $n$-point amplitude of the BAS theory
\begin{align}
\cA_n 
&=
\half \lam   \veps_n
\sum_{P = Q \cup R} f^{\ta_n\tb\tc} \tf^{\ta_n' \tb' \tc'} 
 \phi_Q^{\tb\tb'} \phi_R^{\tc\tc'} \,,
\qquad\qquad P = 12\cdots (n-1) \,.
\end{align}

We now determine how this amplitude transforms under 
the color-factor symmetry associated with scalar $n$.
The color-factor symmetry acts only on the $U(N)$ structure constants
$f^{\ta\tb\tc}$, 
with the $U({\tN})$ structure constants $\tf^{\ta'\tb'\tc'}$ 
behaving  as spectators.\footnote{Naturally, one 
could alternatively define color-factor shifts 
that act on $\tf^{\ta'\tb'\tc'}$.}
From \eqn{cfs} we have 
\begin{align}
\delta_n \, \cA_n 
&=  \half \lam \veps_n   \tf^{\ta_n' \tb' \tc'}  
\sum_{P = B \cup C}
(\delta_n f^{\ta_n\tb\tc}) \phi_B^{\tb\tb'} \phi_C^{\tc\tc'}
\nn\\
&=  \half \lam \alpha_n \veps_n  \tf^{\ta_n' \tb' \tc'}  
\sum_{P = B \cup C} 
\delta^{\tb\tc} (k_B^2 - k_C^2) \phi_B^{\tb\tb'} \phi_C^{\tc\tc'} \,.
\end{align}

Since the sum over divisions of $P$ into words $B$ and $C$
is symmetric under $B \leftrightarrow C$, 
we may relabel
$B, \tb, \tb'  \leftrightarrow C, \tc, \tc'$ 
in the first term
\begin{align}
\tf^{\ta_n' \tb' \tc'}  
\sum_{P = B \cup C} 
\delta^{\tb\tc} k_B^2  \phi_B^{\tb\tb'} \phi_C^{\tc\tc'}
= 
\tf^{\ta_n' \tc' \tb'}  
\sum_{P = B \cup C} 
\delta^{\tc\tb} k_C^2  \phi_C^{\tc\tc'} \phi_B^{\tb\tb'}
\end{align}

so that using 
$ \tf^{\ta_n' \tc' \tb'}  = - \tf^{\ta_n' \tb' \tc'}  $
we have
\begin{align}
\delta_n \, \cA_n 
&=  - \lam \alpha_n \veps_n  \tf^{\ta_n' \tb' \tc'}  
\sum_{P = B \cup C}  
\delta^{\tb\tc} 
\phi_B^{\tb\tb'} k_C^2 \phi_C^{\tc\tc'} \,.
\end{align}

Now we again use \eqn{biadjointrecursion} to obtain
\begin{align}
\delta_n \, \cA_n 
&=  -\half \lam^2 \alpha_n \veps_n  
\tf^{\ta_n' \tb' \tc'}  
\tf^{\tc' \td' \te'}  
\left( f^{\tb \td \te}  
\sum_{P = B \cup D \cup E }  
\phi_B^{\tb\tb'} 
\phi_D^{\td\td'} 
\phi_E^{\te\te'}  \right) \,.
\end{align}

Using the invariance of the sum over  $P= B \cup D \cup E $  
under any permutation of the words $B$, $D$, and $E$,
we observe that the term in parentheses is invariant under cyclic
permutations 
\begin{align}
B, \tb, \tb'   \to D, \td, \td'  \to E, \te, \te'  \to B, \tb, \tb'
\nn
\end{align}

so that we may cyclically symmetrize
$\tf^{\ta_n' \tb' \tc'}  \tf^{\tc' \td' \te'}  $ to obtain
\begin{align}
\delta_n \, \cA_n 
&=  -\fr{1}{6} \lam^2 \alpha_n \veps_n  
\left( 
\tf^{\ta_n' \tb' \tc'}  \tf^{\tc' \td' \te'}  
+\tf^{\ta_n' \td' \tc'}  \tf^{\tc' \te' \tb'}  
+\tf^{\ta_n' \te' \tc'}  \tf^{\tc' \tb' \td'}  
\right)
\left( f^{\tb \td \te}  
\sum_{P = B \cup D \cup E }  
\phi_B^{\tb\tb'} 
\phi_D^{\td\td'} 
\phi_E^{\te\te'}  \right) \,.
\end{align}

Since the term in the left parenthesis vanishes by the Jacobi identity,
the $n$-point amplitude is invariant under the color-factor
symmetry associated with scalar $n$.
Since the amplitude is Bose symmetric, it is invariant under
the color-factor symmetry associated with any of the external fields
\begin{align}
\delta_a \,  \cA_n=0
\end{align}
as was previously established using the cubic vertex 
expansion \cite{Brown:2016mrh}.

\subsection{Yang-Mills theory}

To prove that the tree-level $n$-gluon amplitude 
is invariant under color-factor shifts,
we begin by combining \eqn{ngluon} with \eqns{defAi}{AeomP} to obtain 
\begin{align}
\cA_n =  g \veps_n^{\mu} 
\sum_{P=Q \cup R} f^{\ta_n\tb\tc} A_Q^{\nu\,\tb} G_R^{\nu\mu\,\tc}  \,,
\qquad\qquad P = 12\cdots (n-1) \,.
\end{align}

The color-factor symmetry associated with gluon $n$
acts only on the explicit factor 
$ f^{\ta_n\tb\tc} $ in the equation above, giving
\begin{align}
\delta_n \, \cA_n 
&=  g \veps_n^{\mu} 
\sum_{P=Q \cup R}
(\delta_n f^{\ta_n\tb\tc})
 A_Q^{\nu\,\tb} G_R^{\nu\mu\,\tc} 
\nn\\
&=  g \alpha_n \veps_n^{\mu} 
\sum_{P=Q \cup R} \delta^{\tb\tc} 
(k_Q^2-k_R^2) A_Q^{\nu\,\tb} G_R^{\nu\mu\,\tc} 
\nn\\
&=  g \alpha_n \veps_n^{\mu} 
\sum_{P=Q \cup R}  \left[
(k_Q^2 A_Q^{\lam\,\tc})  G_R^{\lam\mu\,\tc} 
-A_Q^{\nu\,\ta} (k_R^2 G_R^{\nu\mu\,\ta} )
\right] \,.
\label{deltaAn}
\end{align}

Our goal is to show that the right hand side 
of this equation vanishes,
so we must first compute
\begin{align}
S_P^\mu \equiv \sum_{P=Q \cup R}  \left[
(k_Q^2 A_Q^{\lam\,\tc})  G_R^{\lam\mu\,\tc} 
-A_Q^{\nu\,\ta} (k_R^2 G_R^{\nu\mu\,\ta} )
\right]
\label{defS}
\end{align}

which unfortunately is a bit more complicated than the biadjoint scalar case.
First we use \eqns{AeomP}{GeomP} to find
\begin{align}
S_P^\mu 
&=
g f^{\ta\tb\tc}
\sum_{P=A\cup B\cup C}
\left[ 
\left( A_A^{\nu\,\ta} G_B^{\nu \lam \,\tb} \right) G_C^{\lam\mu\,\tc}
- A_A^{\nu\,\ta} \left(
2k_{BC}^\nu A_B^{\lam \,\tb} G_C^{\lam\mu\,\tc}
-k_{BC}^\mu A_B^{\lam \,\tb} G_C^{\lam\nu\,\tc}
-k_{BC}^2 A_B^{\nu \,\tb} A_C^{\mu\,\tc} \right)
\right]
\nn\\
&=
g f^{\ta\tb\tc} \sum_{P=A\cup B\cup C}
A_A^{\nu\,\ta} 
\left[
\left( G_B^{\nu \lam \,\tb} - 2k_B^\nu A_B^{\lam \,\tb} 
- 2k_C^\nu A_B^{\lam \,\tb} \right)  G_C^{\lam\mu\,\tc}
+k_{BC}^\mu A_B^{\lam \,\tb} G_C^{\lam\nu\,\tc}
+k_{BC}^2 A_B^{\nu \,\tb} A_C^{\mu\,\tc}
\right]
\end{align}

where $k^\mu_{BC} = k^\mu_{B} + k^\mu_{C} $.
We use \eqn{defGP} to reexpress this as 
\begin{align}
S_P^\mu
&=
g f^{\ta\tb\tc}
\sum_{P=A\cup B\cup C}
A_A^{\nu\,\ta} 
\Big[
\left( -k_B^\lam A_B^{\nu\,\tb} - 2 k_C^\nu A_B^{\lam\,\tb} - H_B^{\nu\lam\,\tb}\right)
\bigl( 2 k_C^\lam A_C^{\mu\,\tc} - k_C^\mu A_C^{\lam\,\tc} - H_C^{\lam\mu\,\tc} \bigr)
\nn \\
&
\qquad 
+(k_C^\mu + k_B^\mu)  A_B^{\lam \,\tb} 
\left( 2 k_C^\lam A_C^{\nu\,\tc} - k_C^\nu A_C^{\lam\,\tc} - H_C^{\lam\nu\,\tc}\right)
+\left(k_C^2 + 2 k_B \cdot k_C + k_B^2\right)  A_B^{\nu \,\tb} A_C^{\mu\,\tc}
\Big]\,.
\label{expandS}
\end{align}

\Eqn{expandS} can be split into two contributions
\begin{align}
S_P^\mu 
&
= S_{P,1}^\mu + S_{P,2}^\mu \,,
\\[3mm]
S_{P,1}^\mu 
&
= g f^{\ta\tb\tc}
\sum_{P=A\cup B\cup C}
\big[ 
  \left(k_B \cdot A_C^\tc \ A_A^\ta \cdot A_B^\tb
+ 2 k_C \cdot A_A^\ta \ A_B^\tb \cdot A_C^\tc
+ A_A^{\nu\,\ta} H_B^{\nu\lam\,\tb} A_C^{\lam\,\tc} \right) k_C^\mu 
\nn\\
&\quad \qquad \qquad\qquad 
+ \left( 2 k_C \cdot A_B^\tb \ A_A^\ta \cdot A_C^\tc 
- k_C \cdot A_A^\ta \ A_B^\tb \cdot A_C^\tc 
- A_A^{\nu\,\ta} A_B^{\lam \,\tb} H_C^{\lam\nu\,\tc} 
\right) k^\mu_C 
\nn\\[3mm]
&\quad\qquad \qquad\qquad 
+ \left( 2 k_C \cdot A_B^\tb \ A_A^\ta \cdot A_C^\tc 
- k_C \cdot A_A^\ta \ A_B^\tb \cdot A_C^\tc 
- A_A^{\nu\,\ta} A_B^{\lam \,\tb} H_C^{\lam\nu\,\tc} 
\right) k^\mu_B
\big] \,,
\label{S1}
\\[3mm]
S_{P,2}^\mu &= 
g f^{\ta\tb\tc} \sum_{P=A\cup B\cup C}
\left[
- 4 k_C \cdot A_A^\ta \ k_C \cdot A_B^\tb
+ k_C^2  \ A_A^\ta \cdot A_B^\tb 
\right]  A_C^{\mu\,\tc}
\nn\\
&
\quad + g f^{\ta\tb\tc} \sum_{P=A\cup B\cup C}
\left[
-2 A_A^{\nu\,\ta} H_B^{\nu\lam\,\tb} k_C^\lam
+ A_A^{\nu\,\ta} (k_B^2  A_B^{\nu\,\tb})
\right]  A_C^{\mu\,\tc}
\nn\\
&
\quad
+ g f^{\ta\tb\tc}
\sum_{P=A\cup B\cup C}
A_A^{\nu\,\ta} 
\left(  k_B^\lam A_B^{\nu\,\tb} + 2 k_C^\nu A_B^{\lam\,\tb} + H_B^{\nu\lam\,\tb}\right)
H_C^{\lam\mu\,\tc} 
\label{S2}
\end{align}
where $S^\mu_{P,1}$ contains the terms in which the 
free index $\mu$ labels a momentum $k$
and $S^\mu_{P,2}$ contains those in which it labels a field $A$ or $H$.
\para

First we examine $S^\mu_{P,1}$. 
Relabelling $B, \tb \leftrightarrow C, \tc$ 
in the last line (the $k_B^\mu$ term) of \eqn{S1} 
and using $f^{\ta\tc\tb} = - f^{\ta\tb\tc}$,
we obtain two terms
\begin{align}
S_{P,1}^\mu &=  S_{P,1a}^\mu + S_{P,1b}^\mu\,,
\\[3mm]
S_{P,1a}^\mu 
&= 
g f^{\ta\tb\tc}
\sum_{P=A\cup B\cup C}
\big[ 
 -k_B \cdot A_C^\tc \ A_A^\ta \cdot A_B^\tb
+  k_C \cdot A_A^\ta \ A_B^\tb \cdot A_C^\tc
\nn\\
&
\qquad\qquad
\qquad\qquad
+ 2 k_C \cdot A_B^\tb \ A_A^\ta \cdot A_C^\tc 
+ k_B \cdot A_A^\ta \ A_B^\tb \cdot A_C^\tc \big] k^\mu_C \,,
\label{S1a}\\[3mm]
 S_{P,1b}^\mu
&= 
g f^{\ta\tb\tc}
\sum_{P=A\cup B\cup C}
A_A^{\nu\,\ta} 
\left[
 H_B^{\nu\lam\,\tb} A_C^{\lam\,\tc} 
+A_C^{\lam \,\tc} H_B^{\lam\nu\,\tb} 
-A_B^{\lam \,\tb} H_C^{\lam\nu\,\tc} \right] k_C^\mu \,.
\label{S1b}
\end{align}

For $S^\mu_{P,1a}$, 
we relabel $A, \ta \leftrightarrow B, \tb$ 
in the first two terms of \eqn{S1a} 
and use  $f^{\tb\ta\tc} = - f^{\ta\tb\tc}$
to obtain
\begin{align}
S_{P,1a}^\mu &= 
g f^{\ta\tb\tc}
\sum_{P=A\cup B\cup C}
\big[ 
 k_A \cdot A_C^\tc \ A_A^\ta \cdot A_B^\tb
+ k_C \cdot A_B^\tb \ A_A^\ta \cdot A_C^\tc 
+ k_B \cdot A_A^\ta \ A_B^\tb \cdot A_C^\tc \big] k^\mu_C \,.
\label{S1anext}
\end{align}

Then we cyclically relabel the last two terms of \eqn{S1anext}
and use $f^{\tb\tc\ta} = f^{\tc\ta\tb} = f^{\ta\tb\tc}$ 
to obtain
\begin{align}
S_{P,1a}^\mu 
&= 
g f^{\ta\tb\tc}
\sum_{P=A\cup B\cup C}
k_A \cdot A_C^\tc \ A_A^\ta \cdot A_B^\tb 
\ 
(k^\mu_C + k^\mu_A+ k^\mu_B)
\nn\\
&= 
g f^{\ta\tb\tc}
k_P^\mu 
\sum_{P=A\cup B\cup C}
 k_A \cdot A_C^\tc \ A_A^\ta \cdot A_B^\tb
\label{S1afinal}
\end{align}
where we have used $k^\mu_A + k^\mu_B+ k^\mu_C = k^\mu_P $.
\para

Next, we turn to $S^\mu_{P,1b}$, observing that the 
first two terms in \eqn{S1b} cancel 
(since $H_B^{\nu\lam\,\tb} = -H_B^{\lam\nu\,\tb} $), 
leaving 
\begin{align}
S_{P,1b}^\mu &= 
- g f^{\ta\tb\te}
\sum_{P=A\cup B\cup E}
A_A^{\nu\,\ta} 
A_B^{\lam \,\tb} H_E^{\lam\nu\,\te} k_E^\mu \,.
\end{align}

Using \eqn{defHP} with $E = D\cup C$ we have
\begin{align}
S_{P,1b}^\mu &= 
g^2
f^{\ta\tb\te}
f^{\te\tc\td}
\sum_{P=A\cup B\cup C \cup D}
A_A^\ta \cdot A_C^\tc
\ A_B^\tb \cdot A_D^\td 
\ (k_C^\mu +k_D^\mu) \,.
\end{align}

Since 
$f^{\ta\tb\te} f^{\te\tc\td}
A_A^\ta \cdot A_C^\tc \ A_B^\tb \cdot A_D^\td $ 
is invariant under 
$(A,  \ta \leftrightarrow C, \tc; B, \tb \leftrightarrow D, \td)$, 
we replace this with
\begin{align}
S_{P,1b}^\mu 
&= 
\half g^2
f^{\ta\tb\te}
f^{\te\tc\td}
\sum_{P=A\cup B\cup C \cup D}
A_A^\ta \cdot A_C^\tc
\ A_B^\tb \cdot A_D^\td 
\ (k_A^\mu +k_B^\mu+ k_C^\mu +k_D^\mu)
\nn\\
&= 
\half g^2  
f^{\ta\tb\te}
f^{\te\tc\td}
k_P^\mu
\sum_{P=A\cup B\cup C \cup D}
A_A^\ta \cdot A_C^\tc
\ A_B^\tb \cdot A_D^\td 
\label{S1bfinal}
\end{align}

using $k_P^\mu = k_A^\mu +k_B^\mu+ k_C^\mu +k_D^\mu$.
Combining 
\eqns{S1afinal}{S1bfinal}, we have
\begin{align}
S_{P,1}^\mu &=
k_P^\mu 
\left[ g f^{\ta\tb\tc}
\sum_{P=A\cup B\cup C}
 k_A \cdot A_C^\tc \ A_A^\ta \cdot A_B^\tb
+
\half g^2  
f^{\ta\tb\te}
f^{\te\tc\td}
\sum_{P=A\cup B\cup C \cup D}
A_A^\ta \cdot A_C^\tc
\ A_B^\tb \cdot A_D^\td 
\right] \,.
\label{S1final}
\end{align}

Now we turn to $S^\mu_{P,2}$.
The two terms on the first line of \eqn{S2} 
vanish using $f^{\ta\tb\tc} = - f^{\tb\ta\tc}$,
leaving 
\begin{align} 
S_{P,2}^\mu &=  S^\mu_{P,2a} + S^\mu_{P,2b}  \,,
\\[3mm]
S^\mu_{P,2a}
&=
g f^{\ta\ti\tc} \sum_{P=A\cup I\cup C}
\left[
-2 A_A^{\nu\,\ta} H_I^{\nu\lam\,\ti} k_C^\lam
+ A_A^{\nu\,\ta} (k_I^2  A_I^{\nu\,\ti})
\right]  A_C^{\mu\,\tc} \,,
\label{S2a}\\
S^\mu_{P,2b}
&=
g f^{\ta\tb\tj}
\sum_{P=A\cup B\cup J}
A_A^{\nu\,\ta} 
\left(  k_B^\lam A_B^{\nu\,\tb} + 2 k_J^\nu A_B^{\lam\,\tb} + H_B^{\nu\lam\,\tb}\right)
H_J^{\lam\mu\,\tj}  \,.
\label{S2b}
\end{align}

For $S^\mu_{P,2a}$,
we use \eqns{AeomP}{defHP} with $I=B\cup D$ in \eqn{S2a} 
to obtain 
\begin{align}
S_{P,2a}^\mu &= 
g^2 f^{\ta\ti\tc} f^{\ti\tb\td}
\sum_{P=A\cup B\cup C \cup D}
\left[
-2 k_C \cdot A_D^\td \ A_A^\ta \cdot A_B^\tb 
+ A_A^{\nu\,\ta}  A_B^{\lam\,\tb} G_D^{\lam\nu\,\td} 
\right]
A_C^{\mu\,\tc}
\nn\\
&= g^2 f^{\tc\ta\ti} f^{\ti\tb\td}
\sum_{P=A\cup B\cup C \cup D}
\big[
-2 k_C \cdot A_D^\td \ A_A^\ta \cdot A_B^\tb 
+ 2 k_D \cdot A_B^\tb \ A_A^\ta \cdot A_D^\td
\nn\\
& 
\quad
\qquad
\qquad
\qquad
\qquad\qquad
- k_D \cdot A_A^\ta \ A_B^\tb \cdot A_D^\td
- A_A^{\nu\,\ta}  A_B^{\lam\,\tb} 
 H_D^{\lam\nu\,\td}
\big]
A_C^{\mu\,\tc}\,.
\label{S2afinal}
\end{align}

For $S^\mu_{P,2b}$,
we use \eqn{defHP} with 
$J = D\cup C$  in \eqn{S2b} to obtain
\begin{align}
S_{P,2b}^\mu &= 
g^2 f^{\ta\tb\tj} f^{\tj\td\tc}
\sum_{P=A\cup B\cup C \cup D}
\big[
k_B \cdot A_D^\td \ A_A^\ta \cdot A_B^\tb
+2 k_D \cdot A_A^\ta \ A_B^\tb \cdot A_D^\td 
\nn\\
& \quad\qquad\qquad\qquad\qquad\qquad 
+ 2 k_C \cdot A_A^\ta \ A_B^\tb \cdot A_D^\td 
+ A_A^{\nu\,\ta} H_B^{\nu\lam\,\tb} A_D^{\lam\,\td}
\big]  A_C^{\mu\,\tc} \,.
\label{S2bnext}
\end{align}

Letting $A, \ta \leftrightarrow D, \td$ in \eqn{S2bnext} 
and using
$ f^{\td\tb\tj} f^{\tj\ta\tc} = f^{\tc\ta\ti} f^{\ti\tb\td} $,
we obtain 
\begin{align}
S_{P,2b}^\mu &= 
g^2 
f^{\tc\ta\ti}
f^{\ti\tb\td}
\sum_{P=A\cup B\cup C \cup D}
\big[
k_B \cdot A_A^\ta \ A_D^\td \cdot A_B^\tb
+ 2 k_A \cdot A_D^\td \  A_B^\tb \cdot A_A^\ta 
\nn\\
& \quad\qquad\qquad\qquad\qquad\qquad 
+ 2 k_C \cdot A_D^\td \ A_B^\tb \cdot A_A^\ta 
+ A_D^{\nu\,\td}    H_B^{\nu\lam\,\tb} A_A^{\lam\,\ta}
\big] 
 A_C^{\mu\,\tc} \,.
\label{S2bfinal}
\end{align}

Recombining \eqns{S2afinal}{S2bfinal}
and symmetrizing on $B, \tb \leftrightarrow D, \td$ we find
\begin{align}
S_{P,2}^\mu &=  S^\mu_{P,2a} + S^\mu_{P,2b} 
= S^\mu_{P,2c} + S^\mu_{P,2d}  \,,
\\[3mm]
S^\mu_{P,2c} 
&= 
g^2 f^{\tc\ta\ti} f^{\ti\tb\td}
\sum_{P=A\cup B\cup C \cup D}
\big[ 
~  k_A \cdot A_D^\td \  A_B^\tb \cdot A_A^\ta 
+ k_B \cdot A_A^\ta \  A_D^\td \cdot A_B^\tb
+ k_D \cdot A_B^\tb \ A_A^\ta \cdot A_D^\td
\nn\\
& \qquad\qquad\qquad 
- k_A \cdot A_B^\tb \  A_D^\td \cdot A_A^\ta 
- k_B \cdot A_D^\td \ A_A^\ta \cdot A_B^\tb
- k_D \cdot A_A^\ta \ A_B^\tb \cdot A_D^\td \big]
 A_C^{\mu\,\tc} \,,
\label{S2c} 
\\[4mm]
S^\mu_{P,2d}
&=
-2 g^2 f^{\tc\ta\ti} f^{\ti\tb\tj}
\sum_{P=A\cup B\cup C \cup J}
 A_A^{\nu\,\ta}  A_B^{\lam\,\tb}  A_C^{\mu\,\tc} H_J^{\lam\nu\,\tj} \,.
\label{S2d}
\end{align}

For $S^\mu_{P,2c}$,
we cyclically relabel $ABD$ in four of the six terms in \eqn{S2c} to obtain
\begin{align}
S^\mu_{P,2c} 
&= 
g^2 \left(
 f^{\tc\ta\ti} f^{\ti\tb\td}
 +f^{\tc\td\ti} f^{\ti\ta\tb}
 +f^{\tc\tb\ti} f^{\ti\td\ta}
\right)  
\sum_{P=A\cup B\cup C \cup D}
\big[ k_A \cdot A_D^\td \  A_B^\tb \cdot A_A^\ta 
- k_A \cdot A_B^\tb \  A_D^\td \cdot A_A^\ta 
\big]
 A_C^{\mu\,\tc}
\end{align}
which vanishes by the Jacobi identity.
\para

For $S^\mu_{P,2d}$,
we use \eqn{defHP} with $J=D\cup E$
in \eqn{S2d} to obtain
\begin{align}
S_{P,2d}^\mu 
&= 
 -2 g^3 f^{\tc\ta\ti} f^{\ti\tb\tj} f^{\tj\td\te}
\sum_{P=A\cup B\cup C \cup D\cup E}
 A_A^\ta \cdot A_E^\te \ A_B^\tb \cdot A_D^\td  \ A_C^{\mu\,\tc}  \,.
\end{align}

Using the symmetries of 
$ A_A^\ta \cdot A_E^\te \ A_B^\tb \cdot A_D^\td $,
we may replace
\begin{align} 
f^{\tc\ta\ti} f^{\ti\tb\tj} f^{\tj\td\te}
 \to 
\fr{1}{8}
\left\{ 
\left[ 
\left( f^{\tc\ta\ti} f^{\ti\tb\tj} f^{\tj\td\te} +
       f^{\tc\tb\ti} f^{\ti\ta\tj} f^{\tj\te\td} \right) 
+ (\ta \leftrightarrow \te) \right]
+ (\tb \leftrightarrow \td) \right\}
\end{align} 
which vanishes identically.\footnote{This 
may most directly be seen by expressing 
$
f^{\tc\ta\ti} f^{\ti\tb\tj} f^{\tj\td\te}
= \Tr \left(  T^\tc [T^\ta, [T^\tb,[T^\td,T^\te]]] \right)
$
and expanding.} 
Hence we also have that $S^\mu_{P,2d}=0$.
\para

In sum, we have shown that $S^\mu_{P,2}=0$, leaving 
$S^\mu_P = S^\mu_{P,1}$ as given in \eqn{S1final}
\begin{align}
S_P^\mu &= 
k_P^\mu 
\left[ g f^{\ta\tb\tc}
\sum_{P=A\cup B\cup C}
 k_A \cdot A_C^\tc \ A_A^\ta \cdot A_B^\tb
+
\half g^2  
f^{\ta\tb\te}
f^{\te\tc\td}
\sum_{P=A\cup B\cup C \cup D}
A_A^\ta \cdot A_C^\tc
\ A_B^\tb \cdot A_D^\td 
\right] \,.
\label{Sfinal}
\end{align}

Note that for $P=123$ 
the first term is given by
$g f^{\ta\tb\tc} (n_{123}^\mu + n_{231}^\mu + n_{312}^\mu)$
as written in \eqn{kinemnum}, and the second term is absent.
\para

Combining \eqn{deltaAn} with \eqn{Sfinal} 
we therefore have that the change in the amplitude
under the color-factor shift associated with gluon $n$ is
\begin{align}
\delta_n \, \cA_n 
&=  g^2 \alpha_n \veps_n^{\mu}  k_P^\mu 
\left[  f^{\ta\tb\tc}
\sum_{P=A\cup B\cup C}
 k_A \cdot A_C^\tc \ A_A^\ta \cdot A_B^\tb
+
\half g  
f^{\ta\tb\te}
f^{\te\tc\td}
\sum_{P=A\cup B\cup C \cup D}
A_A^\ta \cdot A_C^\tc
\ A_B^\tb \cdot A_D^\td 
\right]
\end{align}

where $P=12\cdots (n-1)$.
Momentum conservation $\sum_{i=1}^n p_i^\mu = 0$
implies $k_P^\mu = -k_n^\mu$.
Since $\veps_n \cdot k_n = 0$, 
we have established that the $n$-gluon amplitude is invariant 
under the color-factor shift associated with gluon $n$.
Since the $n$-gluon amplitude is Bose symmetric, 
it is therefore invariant under a color-factor shift 
associated with any of the external gluons
\begin{align}
\delta_a  \, \cA_n = 0
\end{align}

which is what we set out to prove.

\section{Conclusions}
\setcounter{equation}{0}
\label{sec:concl}

We began by reviewing the color-factor symmetry 
of tree-level amplitudes of the BAS and Yang-Mills theories.
This symmetry acts as a momentum-dependent shift on the color factors,
leaving the amplitude invariant.
The BCJ relations follow as a direct consequence of this symmetry.
\para

Tree-level amplitudes can be 
obtained from Berends-Giele currents,
which are computed recursively.
The recursions relation for the currents can be derived 
from the classical equations of motion of the theory 
using the color-dressed perturbiner formalism.
We used these recursion relations, 
together with a variety of group theory relations,
to prove the invariance of tree-level amplitudes 
under a color-factor shift.
This proof is a (somewhat) easier alternative to the 
proof of color-factor symmetry using the radiation
vertex expansion given in ref.~\cite{Brown:2016mrh},
and is amenable to generalization to other theories. 
\para

Cheung and Mangan \cite{Cheung:2021zvb}
have shown that the color-factor symmetry 
of the BAS theory, 
with scalars transforming in the adjoint of $U(N) \times U(\tN)$,
also applies to the equations of motion of the theory, 
and that the color-factor invariance of the equations of motion
associated with $U(N)$ is related to 
the conservation of current of the 
global symmetry of the Lagrangian under the dual group $U(\tN)$.
It would be interesting to find a similar relation
for Yang-Mills and other theories possessing color-kinematic duality.

\section*{Acknowledgments}
This material is based upon work supported by the 
National Science Foundation under Grant No.~PHY21-11943.
The author thanks Ruihao Xiao for pointing out a typo 
in eq.~(4.36) in v1/v2, 
and also for suggesting using the shorter identity
\begin{align}
f^{\tc\ta\ti} f^{\ti\tb\tj} f^{\tj\td\te}
 \to 
\fr{1}{4}
\left[ 
\left( f^{\tc\ta\ti} f^{\ti\tb\tj} f^{\tj\td\te} +
       f^{\tc\tb\ti} f^{\ti\ta\tj} f^{\tj\te\td} \right) 
+ (\ta \leftrightarrow \td, \tb \leftrightarrow \te)  \right]
=0 \,.
\nonumber
\end{align}


\begin{thebibliography}{10}
\ifx\href\asklfhas\newcommand{\href}[2]{#2}\fi
\ifx\arxivref\asklfhas\newcommand{\arxivref}[2]{\href{http://arxiv.org/abs/#1}{#2}}\fi
\ifx\doiref\asklfhas\newcommand{\doiref}[2]{\href{http://dx.doi.org/#1}{#2}}\fi
\raggedright
\small
\parskip 0pt

\bibitem{Bern:2019prr}
Z.~Bern, J.~J.~Carrasco, M.~Chiodaroli, H.~Johansson and R.~Roiban,
\textit{``{The Duality Between Color and Kinematics and its Applications}''},
\texttt{\arxivref{1909.01358}{arxiv:1909.01358}}.

\bibitem{Bern:2008qj}
Z.~Bern, J.~J.~M.~Carrasco and H.~Johansson,
\textit{``{New Relations for Gauge-Theory Amplitudes}''},
\textsf{\doiref{10.1103/PhysRevD.78.085011}{Phys.~Rev.~D78,~085011~(2008)}},
\texttt{\arxivref{0805.3993}{arxiv:0805.3993}}.

\bibitem{Bern:2010ue}
Z.~Bern, J.~J.~M.~Carrasco and H.~Johansson,
\textit{``{Perturbative Quantum Gravity as a Double Copy of Gauge Theory}''},
\textsf{\doiref{10.1103/PhysRevLett.105.061602}{Phys.Rev.Lett.~105,~061602~(2010)}},
\texttt{\arxivref{1004.0476}{arxiv:1004.0476}}.

\bibitem{Bern:2010yg}
Z.~Bern, T.~Dennen, Y.-t.~Huang and M.~Kiermaier,
\textit{``{Gravity as the Square of Gauge Theory}''},
\textsf{\doiref{10.1103/PhysRevD.82.065003}{Phys.Rev.~D82,~065003~(2010)}},
\texttt{\arxivref{1004.0693}{arxiv:1004.0693}}.

\bibitem{BjerrumBohr:2009rd}
N.~E.~J.~Bjerrum-Bohr, P.~H.~Damgaard and P.~Vanhove,
\textit{``{Minimal Basis for Gauge Theory Amplitudes}''},
\textsf{\doiref{10.1103/PhysRevLett.103.161602}{Phys.~Rev.~Lett.~103,~161602~(2009)}},
\texttt{\arxivref{0907.1425}{arxiv:0907.1425}}.

\bibitem{Stieberger:2009hq}
S.~Stieberger,
\textit{``{Open \& Closed vs. Pure Open String Disk Amplitudes}''},
\texttt{\arxivref{0907.2211}{arxiv:0907.2211}}.

\bibitem{Feng:2010my}
B.~Feng, R.~Huang and Y.~Jia,
\textit{``{Gauge Amplitude Identities by On-shell Recursion Relation in
  S-matrix Program}''},
\textsf{\doiref{10.1016/j.physletb.2010.11.011}{Phys.~Lett.~B695,~350~(2011)}},
\texttt{\arxivref{1004.3417}{arxiv:1004.3417}}.

\bibitem{Chen:2011jxa}
Y.-X.~Chen, Y.-J.~Du and B.~Feng,
\textit{``{A Proof of the Explicit Minimal-basis Expansion of Tree Amplitudes
  in Gauge Field Theory}''},
\textsf{\doiref{10.1007/JHEP02(2011)112}{JHEP~1102,~112~(2011)}},
\texttt{\arxivref{1101.0009}{arxiv:1101.0009}}.

\bibitem{Brown:2016mrh}
R.~W.~Brown and S.~G.~Naculich,
\textit{``{BCJ relations from a new symmetry of gauge-theory amplitudes}''},
\textsf{\doiref{10.1007/JHEP10(2016)130}{JHEP~1610,~130~(2016)}},
\texttt{\arxivref{1608.04387}{arxiv:1608.04387}}.

\bibitem{Brown:2016hck}
R.~W.~Brown and S.~G.~Naculich,
\textit{``{Color-factor symmetry and BCJ relations for QCD amplitudes}''},
\textsf{\doiref{10.1007/JHEP11(2016)060}{JHEP~1611,~060~(2016)}},
\texttt{\arxivref{1608.05291}{arxiv:1608.05291}}.

\bibitem{Brown:2018wss}
R.~W.~Brown and S.~G.~Naculich,
\textit{``{KLT-type relations for QCD and bicolor amplitudes from color-factor
  symmetry}''},
\textsf{\doiref{10.1007/JHEP03(2018)057}{JHEP~1803,~057~(2018)}},
\texttt{\arxivref{1802.01620}{arxiv:1802.01620}}.

\bibitem{Brown:1982xx}
R.~W.~Brown, K.~Kowalski and S.~J.~Brodsky,
\textit{``{Classical Radiation Zeros in Gauge Theory Amplitudes}''},
\textsf{\doiref{10.1103/PhysRevD.28.624}{Phys.Rev.~D28,~624~(1983)}}.

\bibitem{Cachazo:2013iea}
F.~Cachazo, S.~He and E.~Y.~Yuan,
\textit{``{Scattering of Massless Particles: Scalars, Gluons and Gravitons}''},
\textsf{\doiref{10.1007/JHEP07(2014)033}{JHEP~1407,~033~(2014)}},
\texttt{\arxivref{1309.0885}{arxiv:1309.0885}}.

\bibitem{Cheung:2021zvb}
C.~Cheung and J.~Mangan,
\textit{``{Covariant color-kinematics duality}''},
\textsf{\doiref{10.1007/JHEP11(2021)069}{JHEP~2111,~069~(2021)}},
\texttt{\arxivref{2108.02276}{arxiv:2108.02276}}.

\bibitem{Cheung:2022pdk}
C.~Cheung, J.~Parra-Martinez and A.~Sivaramakrishnan,
\textit{``{On-shell correlators and color-kinematics duality in curved
  symmetric spacetimes}''},
\textsf{\doiref{10.1007/JHEP05(2022)027}{JHEP~2205,~027~(2022)}},
\texttt{\arxivref{2201.05147}{arxiv:2201.05147}}.

\bibitem{Berends:1987me}
F.~A.~Berends and W.~T.~Giele,
\textit{``{Recursive Calculations for Processes with n Gluons}''},
\textsf{\doiref{10.1016/0550-3213(88)90442-7}{Nucl.~Phys.~B~306,~759~(1988)}}.

\bibitem{Rosly:1996vr}
A.~A.~Rosly and K.~G.~Selivanov,
\textit{``{On amplitudes in selfdual sector of Yang-Mills theory}''},
\textsf{\doiref{10.1016/S0370-2693(97)00268-2}{Phys.~Lett.~B~399,~135~(1997)}},
\texttt{\arxivref{hep-th/9611101}{hep-th/9611101}}.

\bibitem{Selivanov:1998hn}
K.~G.~Selivanov,
\textit{``{On tree form-factors in (supersymmetric) Yang-Mills theory}''},
\textsf{\doiref{10.1007/s002200050006}{Commun.~Math.~Phys.~208,~671~(2000)}},
\texttt{\arxivref{hep-th/9809046}{hep-th/9809046}}.

\bibitem{Selivanov:1999as}
K.~G.~Selivanov,
\textit{``{Post-classicism in Tree Amplitudes}''},
\texttt{\arxivref{hep-th/9905128}{hep-th/9905128}},
in: \textit{``{34th Rencontres de Moriond: Electroweak Interactions and Unified
  Theories}''},
473--478p.

\bibitem{Lee:2015upy}
S.~Lee, C.~R.~Mafra and O.~Schlotterer,
\textit{``{Non-linear gauge transformations in $D=10$ SYM theory and the BCJ
  duality}''},
\textsf{\doiref{10.1007/JHEP03(2016)090}{JHEP~1603,~090~(2016)}},
\texttt{\arxivref{1510.08843}{arxiv:1510.08843}}.

\bibitem{Mafra:2015vca}
C.~R.~Mafra and O.~Schlotterer,
\textit{``{Berends-Giele recursions and the BCJ duality in superspace and
  components}''},
\textsf{\doiref{10.1007/JHEP03(2016)097}{JHEP~1603,~097~(2016)}},
\texttt{\arxivref{1510.08846}{arxiv:1510.08846}}.

\bibitem{Mafra:2016ltu}
C.~R.~Mafra,
\textit{``{Berends-Giele recursion for double-color-ordered amplitudes}''},
\textsf{\doiref{10.1007/JHEP07(2016)080}{JHEP~1607,~080~(2016)}},
\texttt{\arxivref{1603.09731}{arxiv:1603.09731}}.

\bibitem{Garozzo:2018uzj}
L.~M.~Garozzo, L.~Queimada and O.~Schlotterer,
\textit{``{Berends-Giele currents in Bern-Carrasco-Johansson gauge for $F^3$-
  and $F^4$-deformed Yang-Mills amplitudes}''},
\textsf{\doiref{10.1007/JHEP02(2019)078}{JHEP~1902,~078~(2019)}},
\texttt{\arxivref{1809.08103}{arxiv:1809.08103}}.

\bibitem{Mizera:2018jbh}
S.~Mizera and B.~Skrzypek,
\textit{``{Perturbiner Methods for Effective Field Theories and the Double
  Copy}''},
\textsf{\doiref{10.1007/JHEP10(2018)018}{JHEP~1810,~018~(2018)}},
\texttt{\arxivref{1809.02096}{arxiv:1809.02096}}.

\bibitem{Macrelli:2019afx}
T.~Macrelli, C.~S\"amann and M.~Wolf,
\textit{``{Scattering amplitude recursion relations in
  Batalin-Vilkovisky\textendash{}quantizable theories}''},
\textsf{\doiref{10.1103/PhysRevD.100.045017}{Phys.~Rev.~D~100,~045017~(2019)}},
\texttt{\arxivref{1903.05713}{arxiv:1903.05713}}.

\bibitem{Lopez-Arcos:2019hvg}
C.~Lopez-Arcos and A.~Q.~V\'elez,
\textit{``{L$_{\infty}$-algebras and the perturbiner expansion}''},
\textsf{\doiref{10.1007/JHEP11(2019)010}{JHEP~1911,~010~(2019)}},
\texttt{\arxivref{1907.12154}{arxiv:1907.12154}}.

\bibitem{Gomez:2020vat}
H.~Gomez, R.~L.~Jusinskas, C.~Lopez-Arcos and A.~Q.~Velez,
\textit{``{The $L_{\infty}$ structure of gauge theories with matter}''},
\textsf{\doiref{10.1007/JHEP02(2021)093}{JHEP~2102,~093~(2021)}},
\texttt{\arxivref{2011.09528}{arxiv:2011.09528}}.

\bibitem{Ahmadiniaz:2021fey}
N.~Ahmadiniaz, F.~M.~Balli, C.~Lopez-Arcos, A.~Q.~Velez and C.~Schubert,
\textit{``{Color-kinematics duality from the Bern-Kosower formalism}''},
\textsf{\doiref{10.1103/PhysRevD.104.L041702}{Phys.~Rev.~D~104,~L041702~(2021)}},
\texttt{\arxivref{2105.06745}{arxiv:2105.06745}}.

\bibitem{Ahmadiniaz:2021ayd}
N.~Ahmadiniaz, F.~M.~Balli, O.~Corradini, C.~Lopez-Arcos, A.~Q.~Velez and
  C.~Schubert,
\textit{``{Manifest colour-kinematics duality and double-copy in the
  string-based formalism}''},
\textsf{\doiref{10.1016/j.nuclphysb.2022.115690}{Nucl.~Phys.~B~975,~115690~(2022)}},
\texttt{\arxivref{2110.04853}{arxiv:2110.04853}}.

\bibitem{Gomez:2021shh}
H.~Gomez and R.~L.~Jusinskas,
\textit{``{Multiparticle Solutions to Einstein\textquoteright{}s Equations}''},
\textsf{\doiref{10.1103/PhysRevLett.127.181603}{Phys.~Rev.~Lett.~127,~181603~(2021)}},
\texttt{\arxivref{2106.12584}{arxiv:2106.12584}}.

\bibitem{Armstrong:2022mfr}
C.~Armstrong, H.~Gomez, R.~Lipinski~Jusinskas, A.~Lipstein and J.~Mei,
\textit{``{New recursion relations for tree-level correlators in
  anti\textendash{}de Sitter spacetime}''},
\textsf{\doiref{10.1103/PhysRevD.106.L121701}{Phys.~Rev.~D~106,~L121701~(2022)}},
\texttt{\arxivref{2209.02709}{arxiv:2209.02709}}.

\bibitem{Gomez:2022dzk}
H.~Gomez, R.~Lipinski~Jusinskas, C.~Lopez-Arcos and A.~Quintero~Velez,
\textit{``{One-Loop Off-Shell Amplitudes from Classical Equations of
  Motion}''},
\textsf{\doiref{10.1103/PhysRevLett.130.081601}{Phys.~Rev.~Lett.~130,~081601~(2023)}},
\texttt{\arxivref{2208.02831}{arxiv:2208.02831}}.

\bibitem{Jurco:2019yfd}
B.~Jur\v{c}o, T.~Macrelli, C.~S\"amann and M.~Wolf,
\textit{``{Loop Amplitudes and Quantum Homotopy Algebras}''},
\textsf{\doiref{10.1007/JHEP07(2020)003}{JHEP~2007,~003~(2020)}},
\texttt{\arxivref{1912.06695}{arxiv:1912.06695}}.

\bibitem{Borsten:2021hua}
L.~Borsten, H.~Kim, B.~Jur\v{c}o, T.~Macrelli, C.~Saemann and M.~Wolf,
\textit{``{Double Copy from Homotopy Algebras}''},
\textsf{\doiref{10.1002/prop.202100075}{Fortsch.~Phys.~69,~2100075~(2021)}},
\texttt{\arxivref{2102.11390}{arxiv:2102.11390}}.

\bibitem{Escudero:2022zdz}
V.~G.~Escudero, C.~Lopez-Arcos and A.~Quintero~Velez,
\textit{``{Homotopy double copy and the
  Kawai\textendash{}Lewellen\textendash{}Tye relations for the non-abelian and
  tensor Navier\textendash{}Stokes equations}''},
\textsf{\doiref{10.1063/5.0119508}{J.~Math.~Phys.~64,~2881598~(2023)}},
\texttt{\arxivref{2201.06047}{arxiv:2201.06047}}.

\bibitem{Borsten:2022ouu}
L.~Borsten, H.~Kim, B.~Jurco, T.~Macrelli, C.~Saemann and M.~Wolf,
\textit{``{Colour-kinematics duality, double copy, and homotopy algebras}''},
\textsf{\doiref{10.22323/1.414.0426}{PoS~ICHEP2022,~426~(2022)}},
\texttt{\arxivref{2211.16405}{arxiv:2211.16405}}.

\bibitem{Bern:2010tq}
Z.~Bern, J.~J.~M.~Carrasco, L.~J.~Dixon, H.~Johansson and R.~Roiban,
\textit{``{The Complete Four-Loop Four-Point Amplitude in N=4 Super-Yang-Mills
  Theory}''},
\textsf{\doiref{10.1103/PhysRevD.82.125040}{Phys.~Rev.~D82,~125040~(2010)}},
\texttt{\arxivref{1008.3327}{arxiv:1008.3327}}.

\bibitem{Cvitanovic:1976am}
P.~Cvitanovic,
\textit{``{Group theory for Feynman diagrams in non-Abelian gauge theories}''},
\textsf{\doiref{10.1103/PhysRevD.14.1536}{Phys.~Rev.~D~14,~1536~(1976)}}.

\bibitem{Cvitanovic:1980bu}
P.~Cvitanovic, P.~G.~Lauwers and P.~N.~Scharbach,
\textit{``{Gauge Invariance Structure of Quantum Chromodynamics}''},
\textsf{\doiref{10.1016/0550-3213(81)90098-5}{Nucl.~Phys.~B~186,~165~(1981)}}.

\bibitem{Sondergaard:2011iv}
T.~Sondergaard,
\textit{``{Perturbative Gravity and Gauge Theory Relations: A Review}''},
\textsf{\doiref{10.1155/2012/726030}{Adv.~High~Energy~Phys.~2012,~726030~(2012)}},
\texttt{\arxivref{1106.0033}{arxiv:1106.0033}}.

\bibitem{DelDuca:1999ha}
V.~Del~Duca, A.~Frizzo and F.~Maltoni,
\textit{``{Factorization of tree QCD amplitudes in the high-energy limit and in
  the collinear limit}''},
\textsf{\doiref{10.1016/S0550-3213(99)00657-4}{Nucl.~Phys.~B568,~211~(2000)}},
\texttt{\arxivref{hep-ph/9909464}{hep-ph/9909464}}.

\bibitem{DelDuca:1999rs}
V.~Del~Duca, L.~J.~Dixon and F.~Maltoni,
\textit{``{New color decompositions for gauge amplitudes at tree and loop
  level}''},
\textsf{\doiref{10.1016/S0550-3213(99)00809-3}{Nucl.~Phys.~B571,~51~(2000)}},
\texttt{\arxivref{hep-ph/9910563}{hep-ph/9910563}}.

\bibitem{Melia:2013bta}
T.~Melia,
\textit{``{Dyck words and multiquark primitive amplitudes}''},
\textsf{\doiref{10.1103/PhysRevD.88.014020}{Phys.~Rev.~D88,~014020~(2013)}},
\texttt{\arxivref{1304.7809}{arxiv:1304.7809}}.

\bibitem{Melia:2013epa}
T.~Melia,
\textit{``{Getting more flavor out of one-flavor QCD}''},
\textsf{\doiref{10.1103/PhysRevD.89.074012}{Phys.~Rev.~D89,~074012~(2014)}},
\texttt{\arxivref{1312.0599}{arxiv:1312.0599}}.

\bibitem{Melia:2015ika}
T.~Melia,
\textit{``{Proof of a new colour decomposition for QCD amplitudes}''},
\textsf{\doiref{10.1007/JHEP12(2015)107}{JHEP~1512,~107~(2015)}},
\texttt{\arxivref{1509.03297}{arxiv:1509.03297}}.

\bibitem{Johansson:2015oia}
H.~Johansson and A.~Ochirov,
\textit{``{Color-Kinematics Duality for QCD Amplitudes}''},
\textsf{\doiref{10.1007/JHEP01(2016)170}{JHEP~1601,~170~(2016)}},
\texttt{\arxivref{1507.00332}{arxiv:1507.00332}}.

\bibitem{delaCruz:2015dpa}
L.~de~la~Cruz, A.~Kniss and S.~Weinzierl,
\textit{``{Proof of the fundamental BCJ relations for QCD amplitudes}''},
\textsf{\doiref{10.1007/JHEP09(2015)197}{JHEP~1509,~197~(2015)}},
\texttt{\arxivref{1508.01432}{arxiv:1508.01432}}.

\end{thebibliography}

\end{document}